\documentclass{PoS}
\pdfoutput=1

\usepackage[utf8]{inputenc}
\usepackage{booktabs}

\title{Constraining the gluon PDF at large $x$ with LHC data}
\ShortTitle{Constraining the gluon PDF at large $x$ with LHC data}

\author{\speaker{Emanuele R. Nocera}\thanks{Joint contribution.}\\
       Rudolf Peierls Centre for Theoretical Physics, University of Oxford\\
       1 Keble Road, OX1 3NP Oxford, United Kingdom\\
       E-mail: \email{emanuele.nocera@physics.ox.ac.uk}}

\author{Maria Ubiali$^{*\dag}$\\
       Cavendish Laboratory, HEP Group, University of Cambridge\\
       19 J.J. Thomson Avenue, CB3 0HE Cambridge, United Kingdom\\
       E-mail: \email{ubiali@hep.phy.cam.ac.uk}}

\abstract{The NNPDF collaboration has recently presented NNPDF3.1, 
a new determination of the parton distribution functions (PDFs) of the 
proton including a number of new data, some of which are 
particularly sensitive to the gluon PDF at large $x$.
In this contribution, we supplement NNPDF3.1 with two new analyses.
First, we study the impact of the ATLAS jet data on the gluon PDF including 
different jet rapidity bins. 
Second, we quantitatively compare the constraints provided by 
all measurements sensitive to the large-$x$ gluon PDF, namely inclusive jet 
cross sections, the transverse momentum distributions of the 
$Z$-boson and top-pair rapidity distributions.}

\FullConference{XXV International Workshop on Deep-Inelastic Scattering 
                and Related Subjects\\
		3-7 April 2017\\
		University of Birmingham, UK}

\begin{document}

\paragraph{Introduction.}

In a recent paper~\cite{Ball:2017nwa} the NNPDF Collaboration presented
NNPDF3.1, a new set of parton distribution functions (PDFs)~\cite{Gao:2017yyd} 
which features several improvements with respect to the previous global
analysis, NNPDF3.0~\cite{Ball:2014uwa}, both in terms of methodology and data.
On the methodological side, the charm PDF was parametrised on the same footing
as the other light quark and gluon PDFs.
On the data side, a number of new measurements, especially from Large Hadron
Collider (LHC) experiments, were included.
Among these, some are directly sensitive to the gluon PDF at medium-to-large
values of Bjorken $x$, which allowed us to determine it with a much 
improved precision.
This is a remarkable feature of NNPDF3.1, as a detailed 
knowledge of the gluon PDF at large $x$ is increasingly crucial in order to 
generate precise predictions of both the signal and the backgrounds in searches 
for new massive particles at the LHC.

In this contribution we present two new analyses that were not
included in NNPDF3.1. 
First, we study the stability of the NNPDF3.1 fit upon the inclusion of either 
of the five bins in which the inclusive 7 TeV ATLAS 2011 jet 
data~\cite{Aad:2014vwa} are provided.
We demonstrate that the particular choice made in the default NNDPF3.1 fit,
{\it i.e.} the central rapidity bin, does not affect the ensuing gluon PDF.
Second, we provide a quantitative comparison among the constraints provided
by the three different datasets included in NNDPF3.1 that are 
sensitive to the gluon PDF at large $x$: inclusive jet cross section, 
$Z$-boson transverse momentum distribution, and top rapidity distribution data.
We make explicit the impact of each of these observables by adding the 
corresponding datasets, one at a time, to a baseline dataset that does not 
include any of them.
This is different from what was presented in the NNPDF3.1 paper, in which one 
dataset at a time was removed from the global fit. 

\paragraph{Stability of NNPDF3.1 upon the choice of the jet bin.}

The NNPDF3.1 analysis included for the first time the single-inclusive 
jet cross sections measured in the 2011 run at 7~TeV with $R=0.6$ by ATLAS
and at 2.76~TeV with $R=0.7$ by CMS. 
These were added on top of four measurements already included in NNPDF3.0, 
namely: CDF Run II kT, CMS 2011, 2010 ATLAS 7 TeV and ATLAS 2.76 TeV, 
including correlations to the 7 TeV data (see~\cite{Ball:2017nwa} for the
experimental references). 

Although next-to-next-to-leading order (NNLO) corrections to the inclusive jet 
production cross section are now known~\cite{Currie:2016bfm} (in the 
leading-colour approximation), the exact results are not  
yet available for all jet datasets included in NNPDF3.1.
Therefore, they were included in the NNPDF3.1 NNLO PDF fit using NNLO PDF 
evolution but next-to-leading order (NLO) matrix elements. 
A fully correlated theoretical systematic uncertainty, accounting
for the missing higher order corrections in the matrix element,
was added to the covariance matrix.
We also note that the sign and the size of the NNLO corrections strongly 
depend on the central scale used in the predictions. 
If the jet transverse momentum $p_T$ is taken as the central scale, the 
NNLO/NLO $K$-factors vary between -5\% and +10\% in the range measured at the 
LHC, $100$ GeV$\lesssim p_T\lesssim 2$ TeV~\cite{Currie:2017ctp}.

While no cuts were applied to all jet datasets included in NNPDF3.1,
for the 2011 ATLAS 7 TeV dataset a good agreement between data and theory
was obtained when fitting only the central rapidity bin, $|y_{\rm jet}| < 0.5$.
Concurrently, it was found that achieving a good description of the ATLAS
2011 7 TeV dataset would be impossible, if all five rapidity bins were 
included simultaneously and if all cross-correlations among 
rapidity bins were taken into account accordingly. 
%
%
It is therefore important to demonstrate that the gluon PDF is stable upon the 
choice of any of the other rapidity bins.
In order to investigate on this, we have performed five additional fits, 
with the same theoretical settings of the default NNPDF3.1 NNLO PDF fit, 
in which we have included in turn the second, third, fourth,
fifth and sixth jet rapidity bin ($0.5<|y_{\rm jet}| < 1.0$, 
$1.0<|y_{\rm jet}| < 1.5$, $1.5<|y_{\rm jet}| < 2.0$, 
$2.0<|y_{\rm jet}| < 2.5$ and $2.5<|y_{\rm jet}| < 3.0$ respectively) 
instead of the central bin.

\begin{figure}[!t]
\centering
\includegraphics[scale=0.18,angle=270,clip=true,trim=0 7cm 0 10cm]{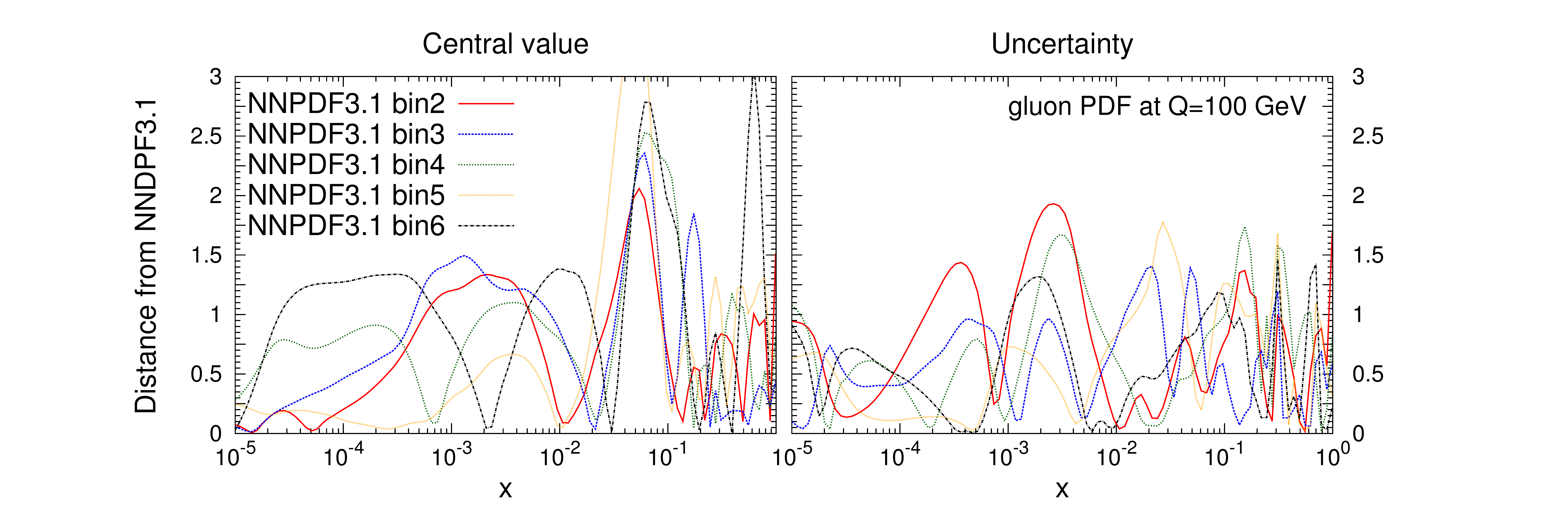}\\
\caption{Distances between the central value (left) and the uncertainty (right) 
  of the gluon PDF at $Q=100$ GeV in the NNPDF3.1 default NNLO fit and in the 
  five variants in which higher jet rapidity bins are included instead of the 
  central bin, see the text for details.}
\label{fig:jetbinsdist}
\end{figure}

\begin{figure}[!t]
\centering
\includegraphics[width=0.47\textwidth]{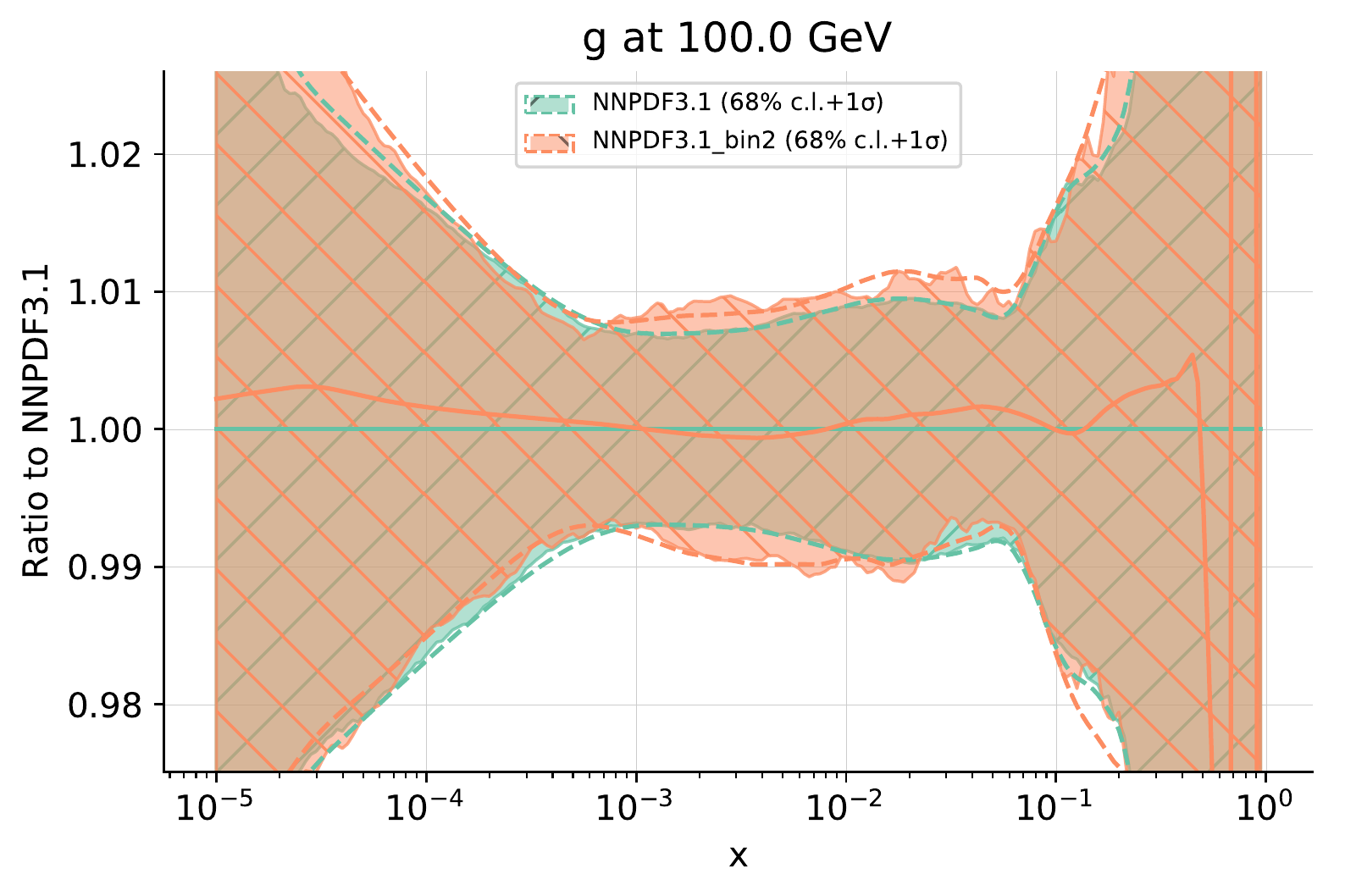}
\includegraphics[width=0.47\textwidth]{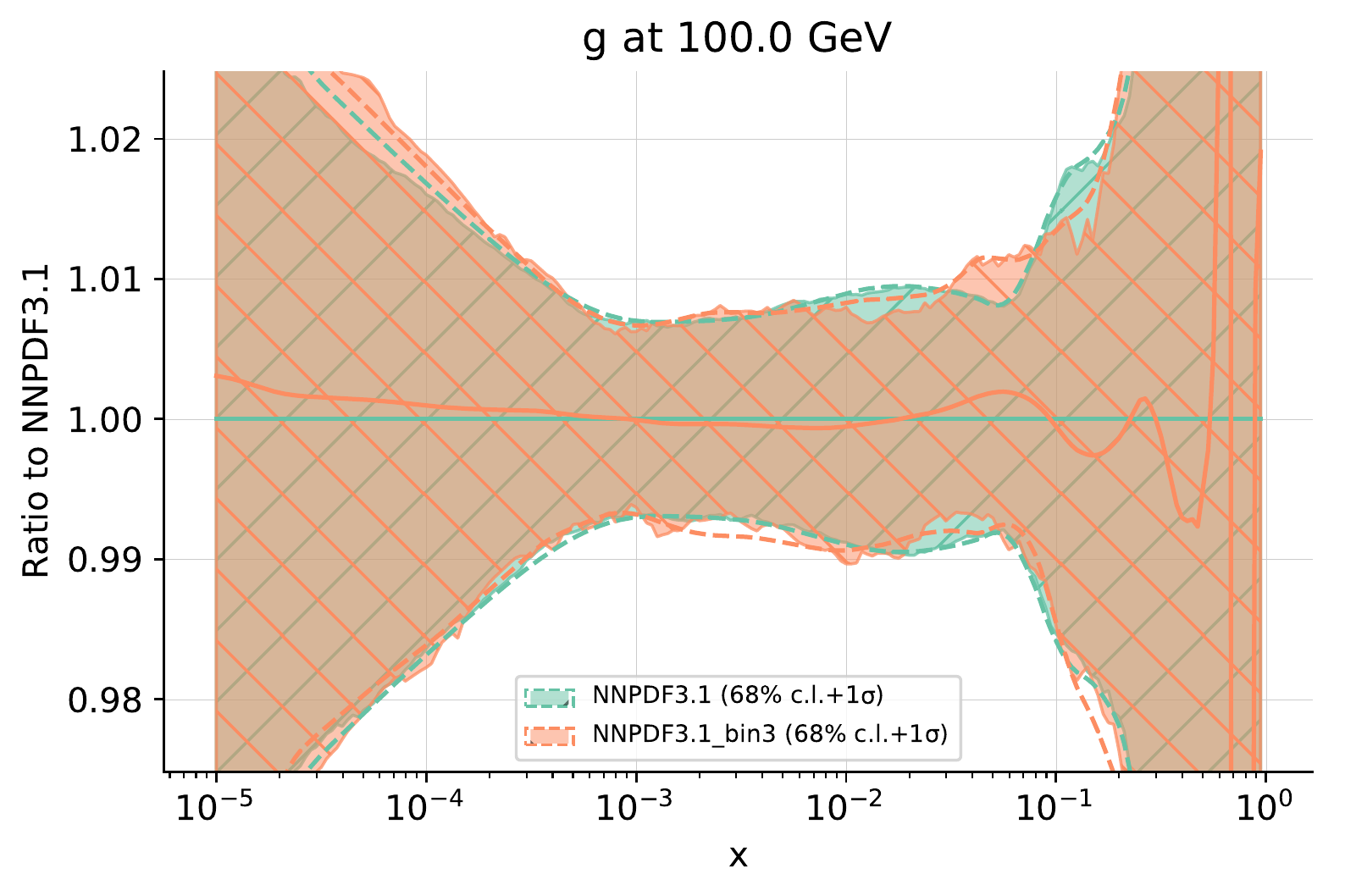}
\caption{Comparison between the NNPDF3.1 default
  NNLO fit and two fits in which the second (left panel) and third
  (right panel) rapidity bin of the ATLAS 2011 7 TeV data are included
  instead of the central bin. The factorisation scale is set to
  $Q=100$ GeV and PDFs are normalised to the NNPDF3.1 set. }
\label{fig:jetbins}
\end{figure}

In Table~\ref{tab:1}, we report the value of the $\chi^2$ per data point,
$\chi^2/N_{\rm dat}$, for the individual 2011 ATLAS 7 TeV data set, before and 
after each of the five variants of the default fit, and for the total data set 
after the fits.
In Fig.~\ref{fig:jetbinsdist}, we show the distance~\cite{Ball:2010de} 
between the central value and the uncertainty of the gluon PDF at $Q=100$ GeV 
in the default NNLO NNPDF3.1 fit and in each of the five variants of the fit 
including higher rapidity bins. 
All sets are made of $N_{\rm rep}=100$ replicas.
Distances of $d\simeq 1$ correspond to statistically equivalent fits, 
while for sets of 100 replicas $d\simeq 10$ corresponds to a difference 
of one sigma in unity of the corresponding variance. 
In Fig.~\ref{fig:jetbins} we compare the NNLO gluon PDF obtained from the 
NNPDF3.1 default fit and the variants in which the second and third rapidity 
bins of the 2011 ATLAS 7 TeV jet data are fitted instead of the central bin.
Very similar plots are found for fits with higher rapidity bins, therefore
they are not shown. 
As is apparent from Table~\ref{tab:1} and 
Figs.~\ref{fig:jetbinsdist}-\ref{fig:jetbins},
the description of each separate bin is equally good, the central values 
of the gluon PDF are well within its uncertainty for each fit
and PDFs are statistically equivalent. 
Therefore, we conclude that the gluon PDF in the NNPDF3.1 set is 
independent of the choice of the ATLAS 2011 jet data bin used in the fit.

\begin{table}[!t]
\centering
\footnotesize
\begin{tabular}{lccc}
\toprule
Fit 
& $\chi^2_{\rm ATLAS}/N_{\rm dat}$ (before fit) 
& $\chi^2_{\rm ATLAS}/N_{\rm dat}$ (after fit) 
& $\chi^2_{\rm tot}/N_{\rm dat}$ (after fit)\\
\midrule
NNPDF3.1 centralbin  & 1.07 & 1.07 & 1.148 \\
NNPDF3.1 bin2        & 1.27 & 1.27 & 1.150 \\
NNPDF3.1 bin3        & 0.95 & 0.93 & 1.151 \\
NNPDF3.1 bin4        & 1.06 & 1.07 & 1.145 \\
NNPDF3.1 bin5        & 0.97 & 0.96 & 1.146 \\
NNPDF3.1 bin6        & 0.73 & 0.67 & 1.145 \\
\bottomrule
\end{tabular}
\caption{The $\chi^2$ per data point, $\chi^2/N_{\rm dat}$, for the 2011 ATLAS
  7 TeV data set, before and after each variant of the default fit, and for 
  the total data set after the fits. The first fit is the default NNPDF3.1 
  NNLO fit with $\alpha_s(M_Z) = 0.118$, in which only the central jet rapidity 
  bin ($|y_{\rm jet}| < 0.5$) is included. 
  In the following fits the central bin is replaced by the second, third, 
  fourth, fifth and sixth rapidity bin respectively.}
\label{tab:1}
\end{table}

\paragraph{Impact of various datasets on the gluon PDF at large $x$.}

On top of the inclusive jet data, the two leading 
observables sensitive to the medium-to-large $x$ gluon PDF,
that were included in NNDPF3.1, are the total rate and differential 
distributions of top-pair production and the transverse momentum distributions 
of the $Z$ boson. 
All of them have been measured by the ATLAS and CMS experiments at the LHC 
with high precision recently.
The corresponding NNLO QCD corrections have been computed 
for top-pair production cross sections at the level of total rates
in~\cite{Czakon:2013goa,Czakon:2012pz,Baernreuther:2012ws}
and of differential cross sections 
in~\cite{Czakon:2015owf,Czakon:2014xsa,Czakon:2016ckf,Czakon:2016dgf};
for the transverse momentum of the $Z$ boson 
in~\cite{Boughezal:2015ded,Ridder:2016nkl,Gehrmann-DeRidder:2016jns}.
%
%
The impact on the gluon PDF of the newly available data sets for each of 
these observables has been studied in detail~\cite{Beneke:2012wb,Czakon:2013tha,
Czakon:2016olj,Boughezal:2017nla}.
They have been shown to provide complementary and compatible constraints onto 
the gluon PDF at medium and large $x$, and that this constraint is competitive 
with that provided by inclusive jet data.

Specifically, the new data sets directly sensitive to the gluon PDF at large 
$x$ that were included in NNPDF3.1 are the following: the $Z$ boson 
$(p_T^Z,y_Z)$ and $(p_T^Z,M_{ll})$ double differential distributions at 8~TeV 
from ATLAS and the $Z$ boson $(p_T^Z,y_Z)$ double differential distribution
at 8~TeV from CMS; the top-pair production normalised $y_t$ 
differential distribution at 8~TeV from ATLAS, the top-pair production 
normalised $y_{t\bar{t}}$ differential distribution at 8~TeV from CMS
and the total inclusive cross sections for top-pair production at 7, 8 and 
13~TeV from ATLAS and CMS (for all references to the experimental papers, 
see~\cite{Ball:2017nwa}).

\begin{figure}[!t]
\centering
\includegraphics[scale=0.18,angle=270,clip=true,trim=0 7cm 0 10cm]{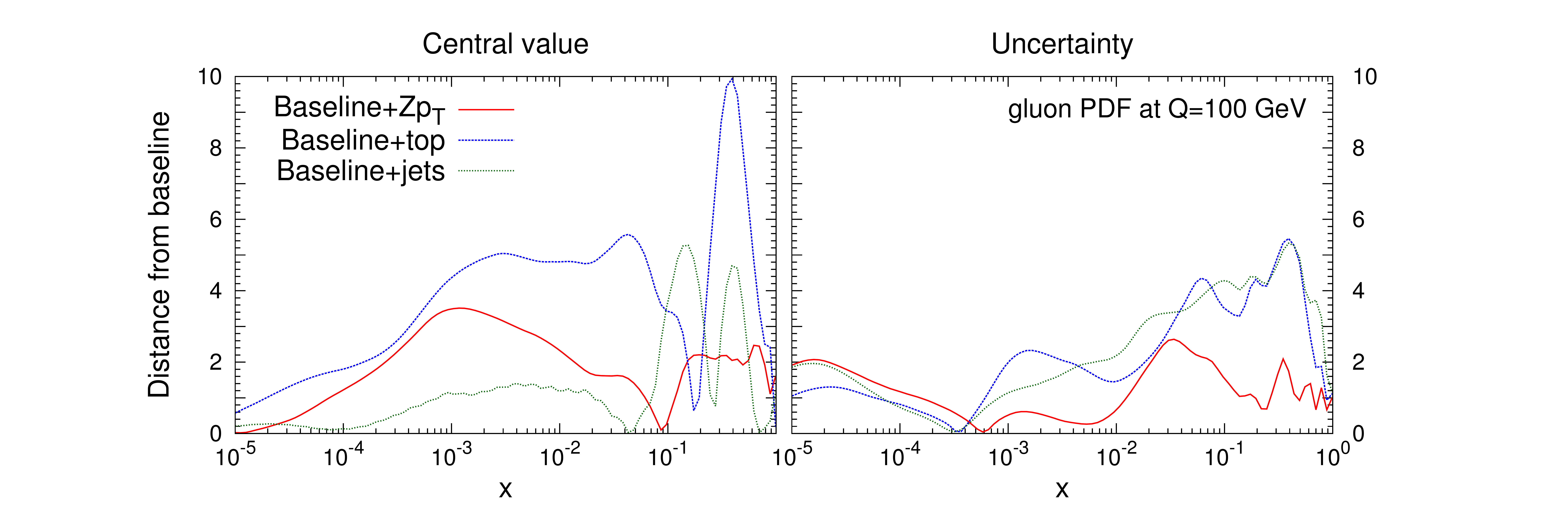}\\
\caption{Distances between the central value (left) and the uncertainty (right)
of the gluon PDF at $Q+100$ GeV in the NNPDF3.1 baseline NNLO fit without
jet, $Z$ $p_T$ and top data and the three variants in which each of these
datasets is added at a time on top of the baseline.}
\label{fig:largexdist}
\end{figure}

\begin{figure}[!p]
\centering
\includegraphics[width=0.47\textwidth]{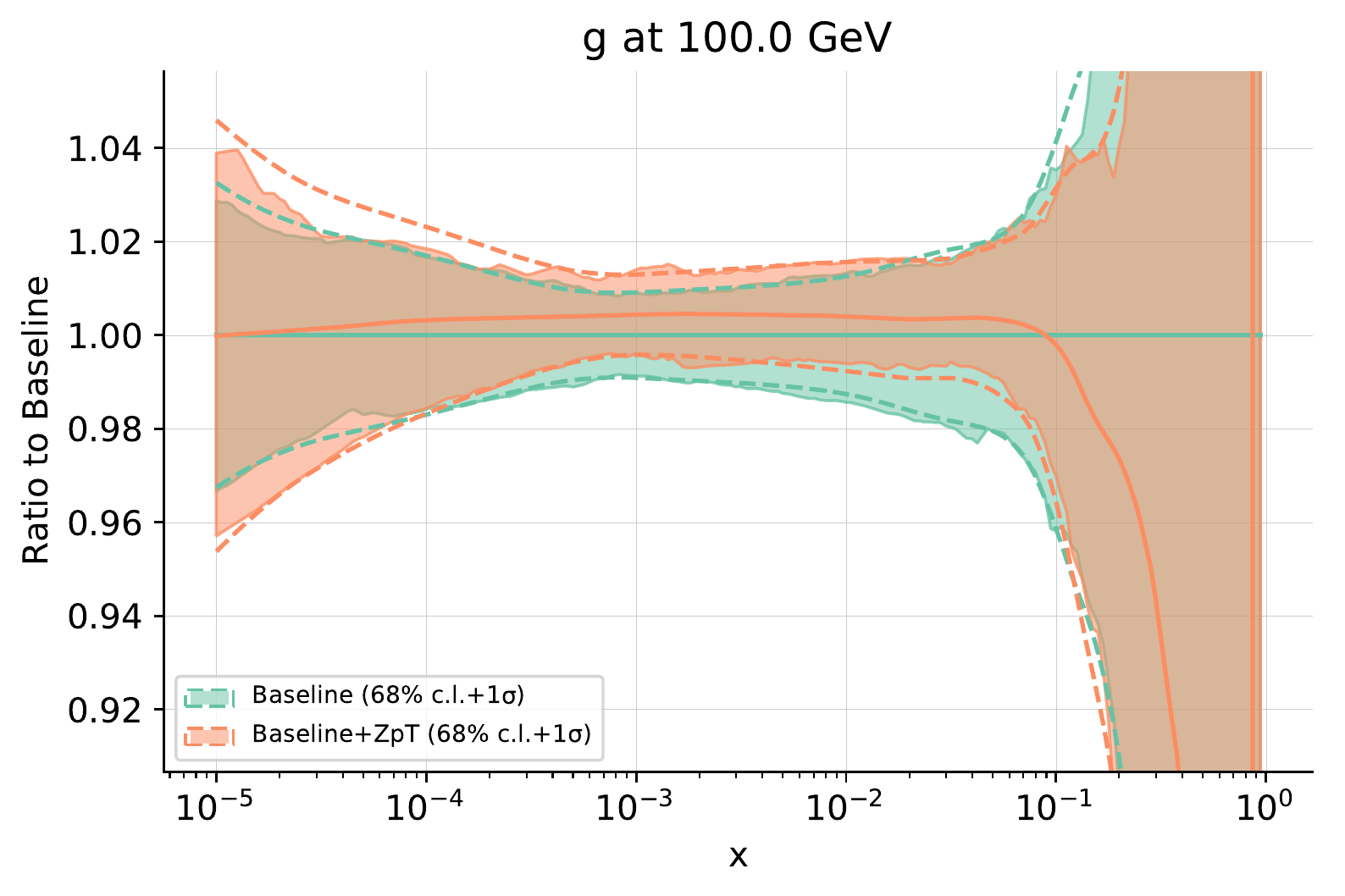}
\includegraphics[width=0.47\textwidth]{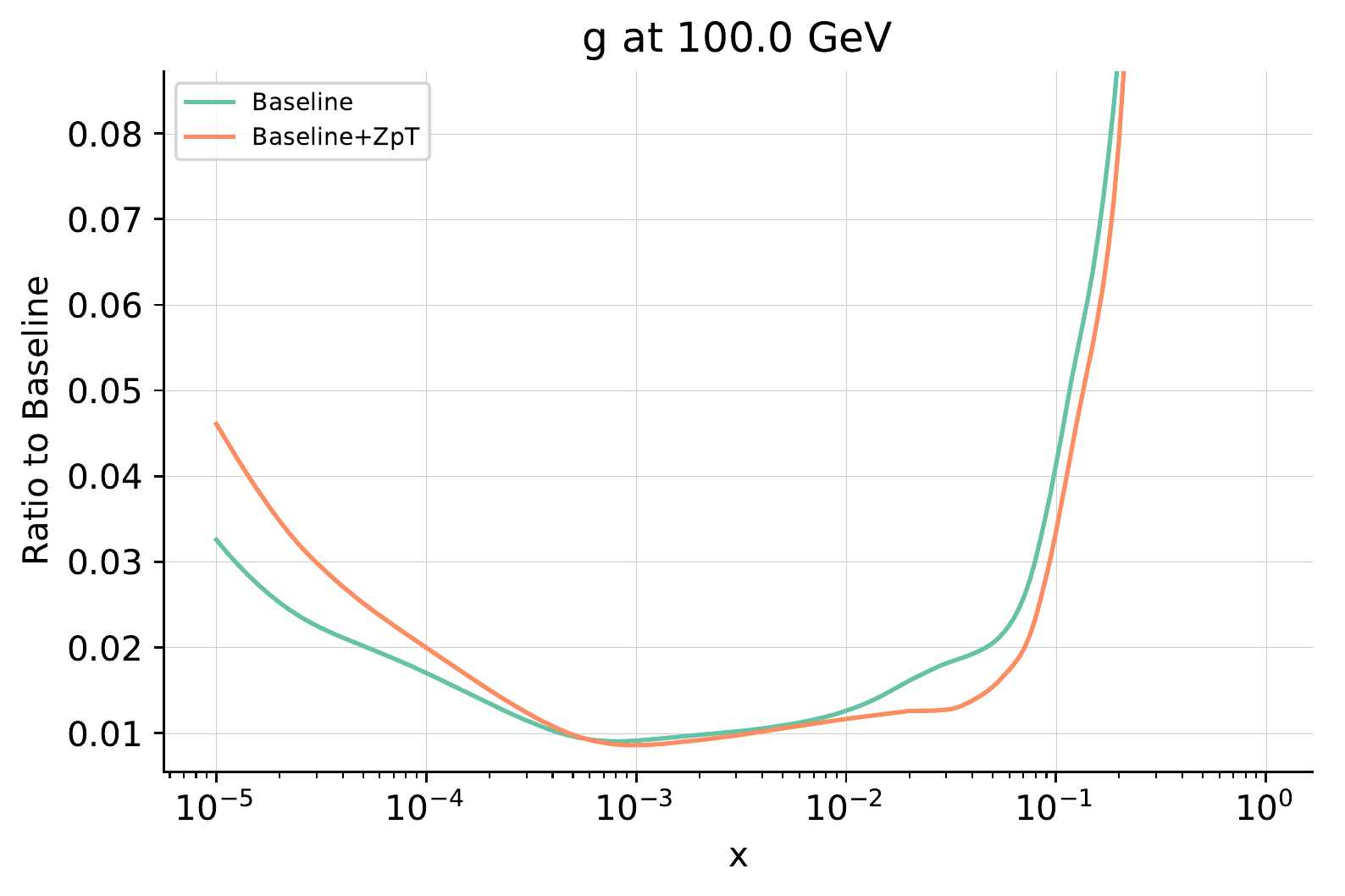}\\
\includegraphics[width=0.47\textwidth]{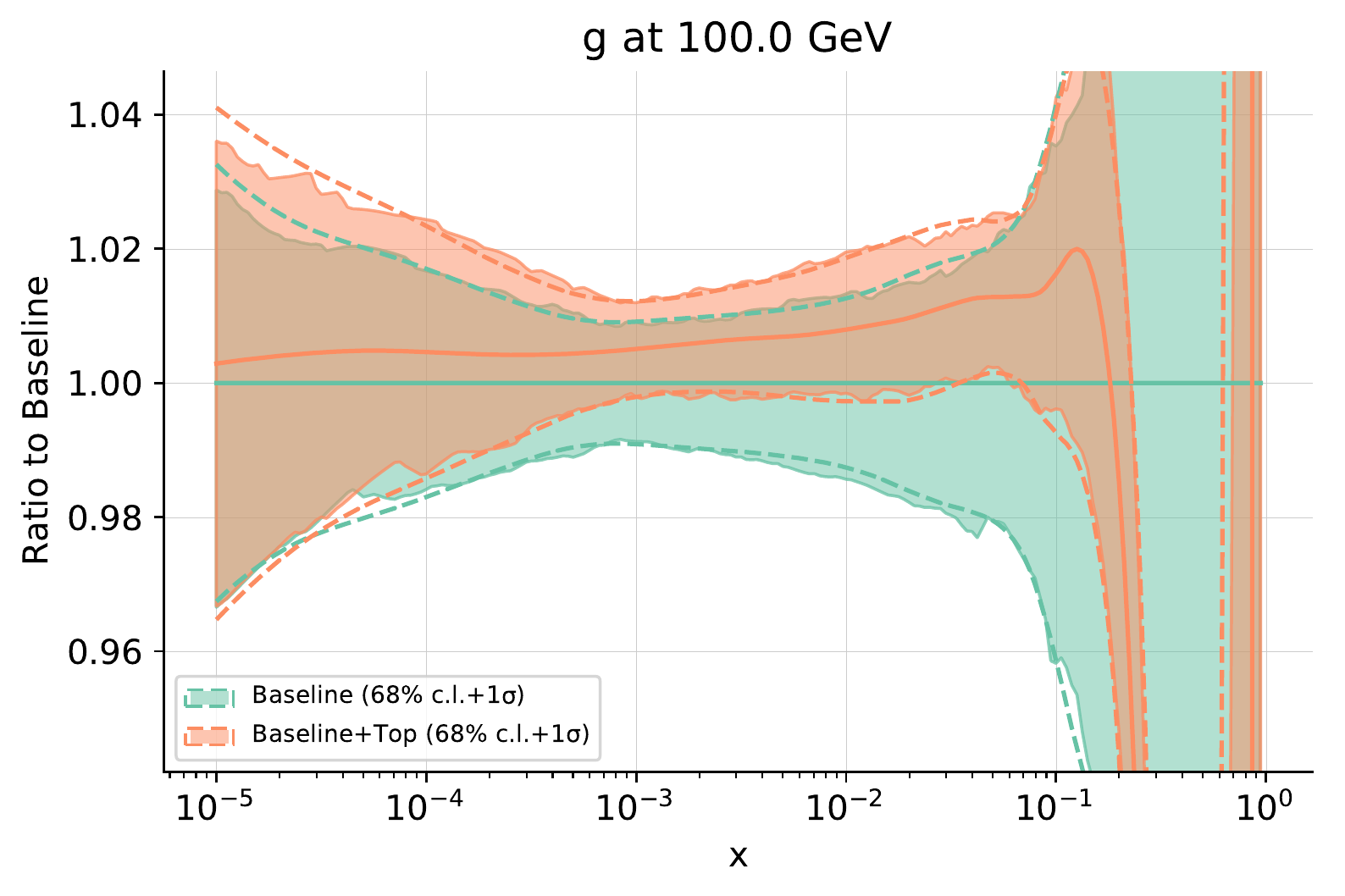}
\includegraphics[width=0.47\textwidth]{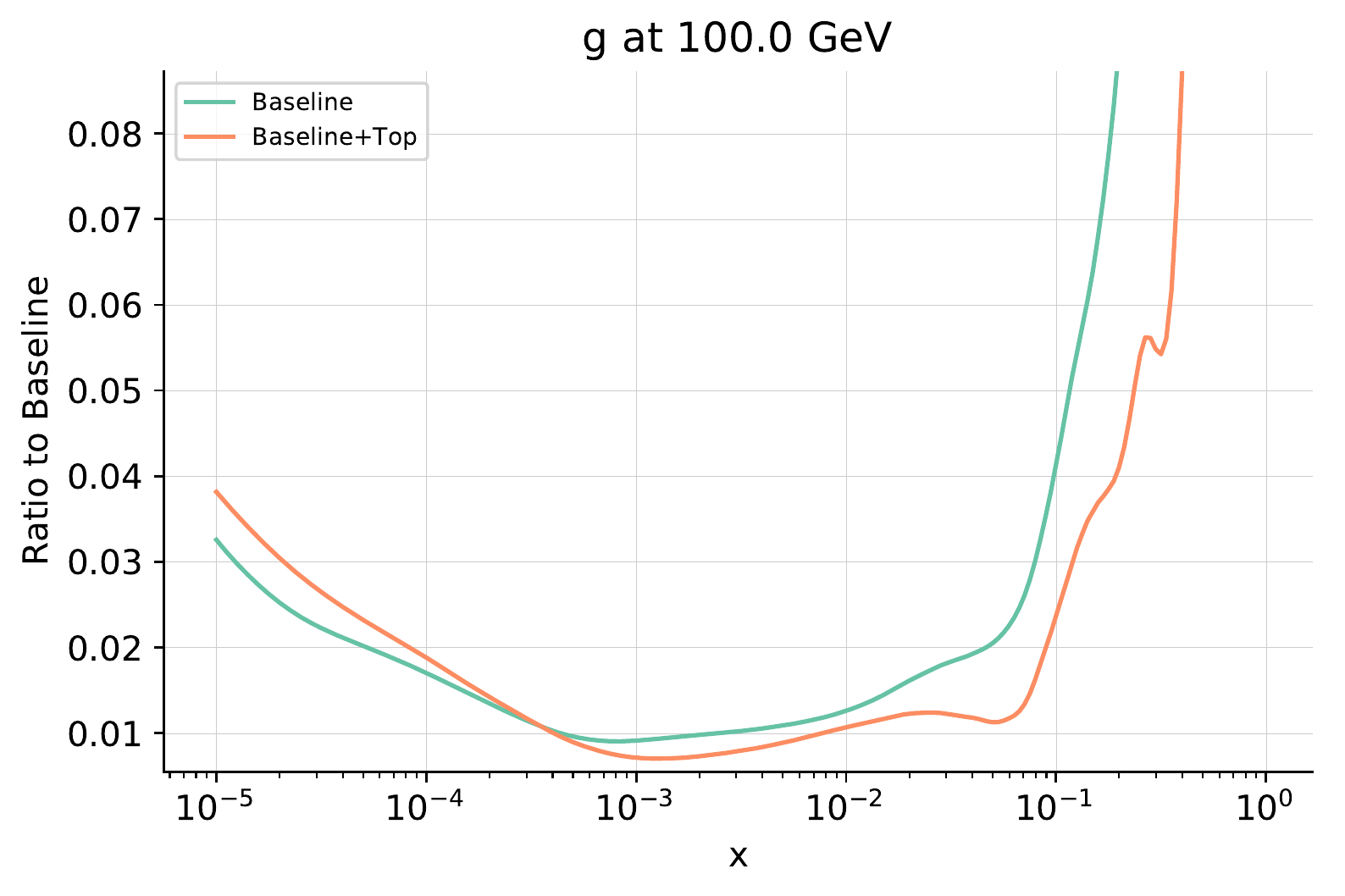}\\
\includegraphics[width=0.47\textwidth]{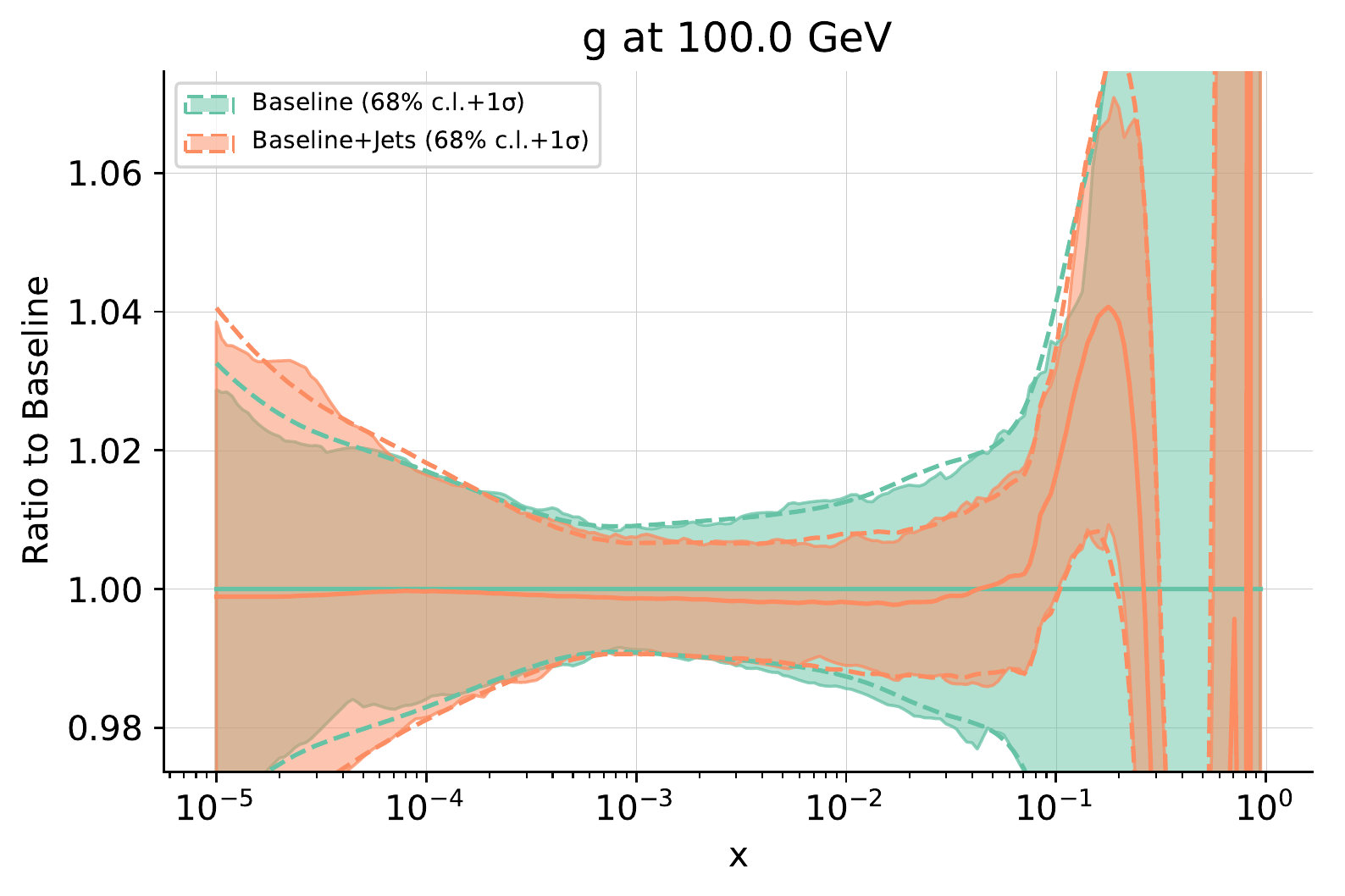}
\includegraphics[width=0.47\textwidth]{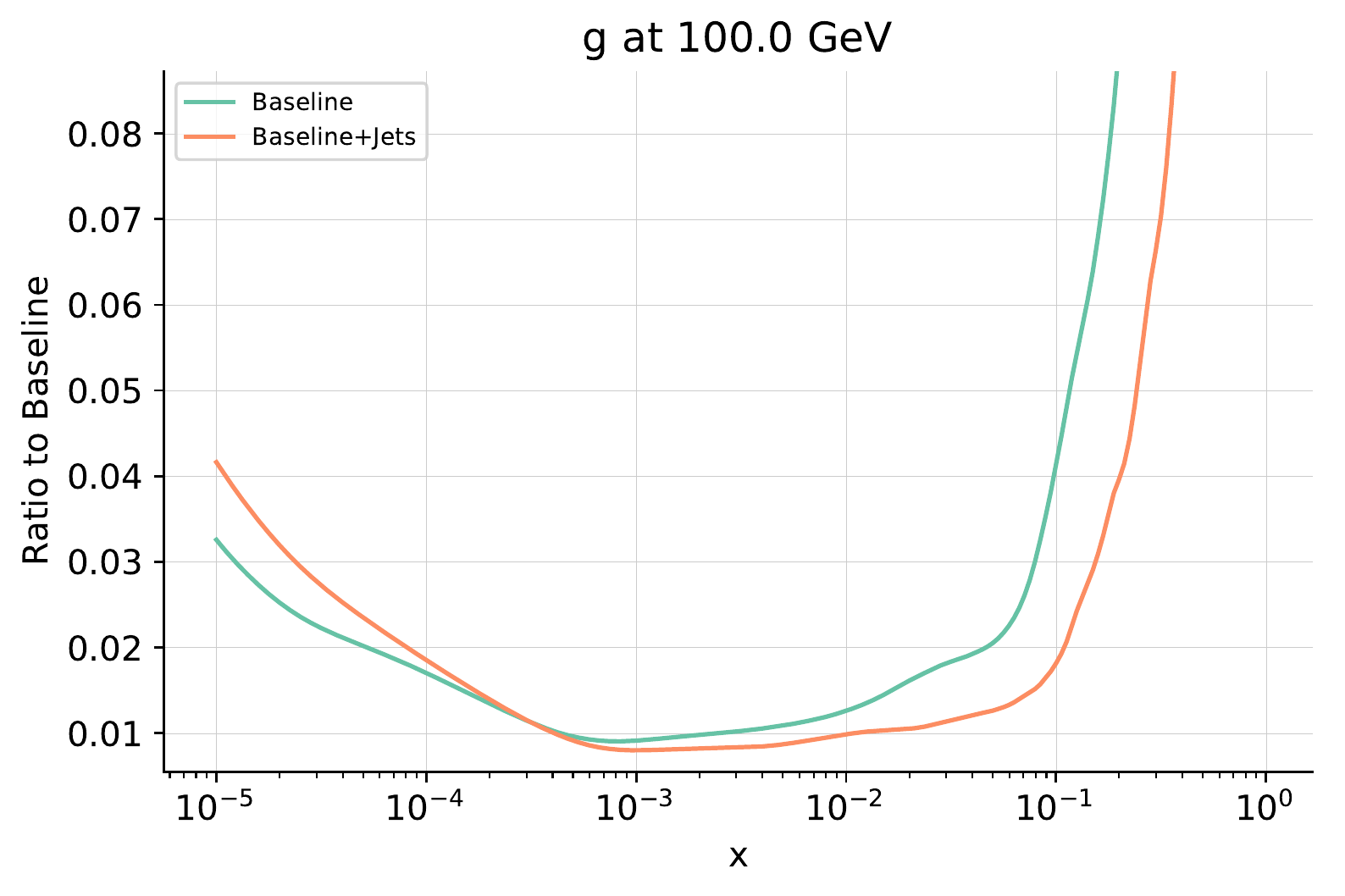}\\
\includegraphics[width=0.47\textwidth]{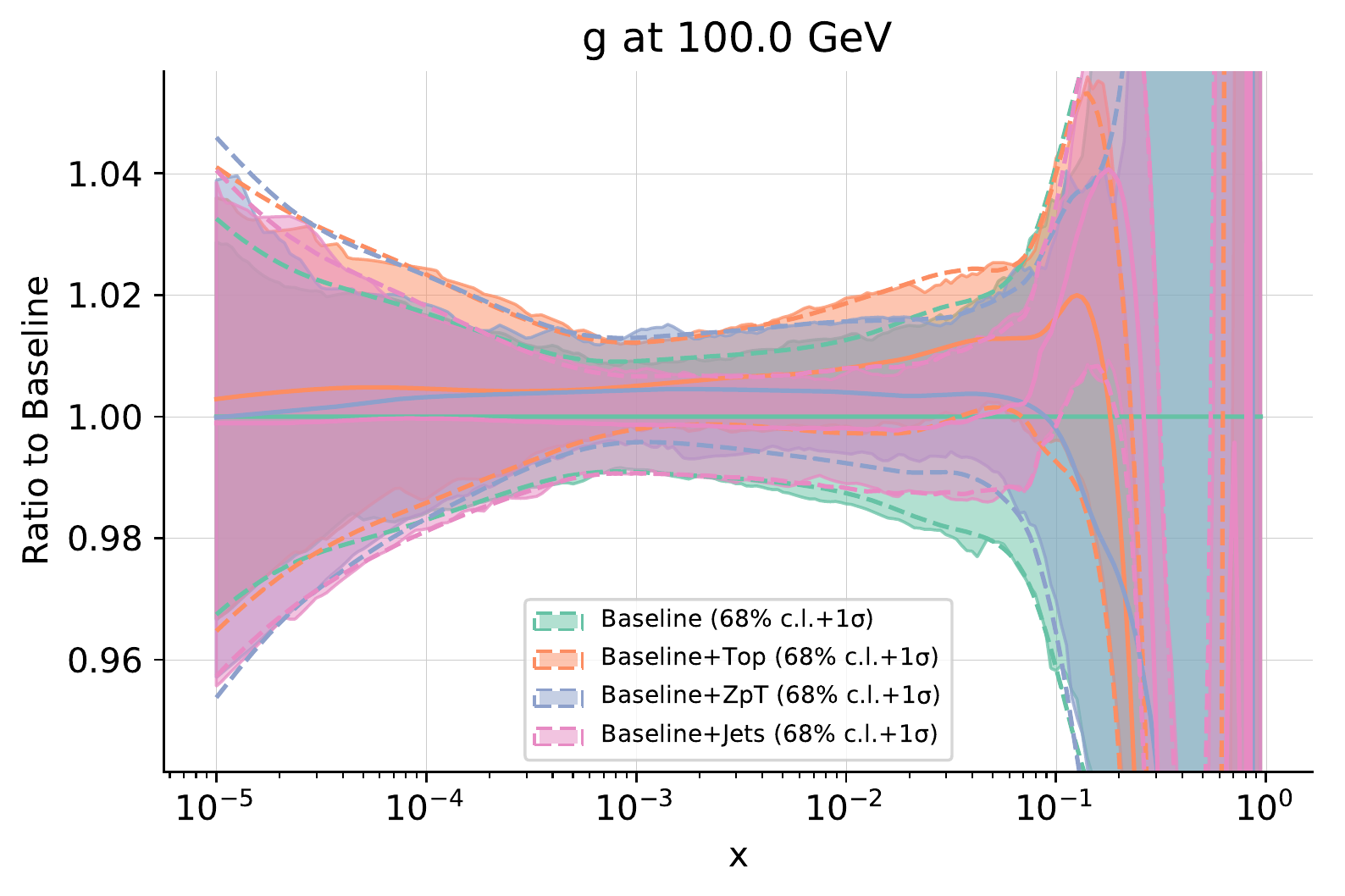}
\includegraphics[width=0.47\textwidth]{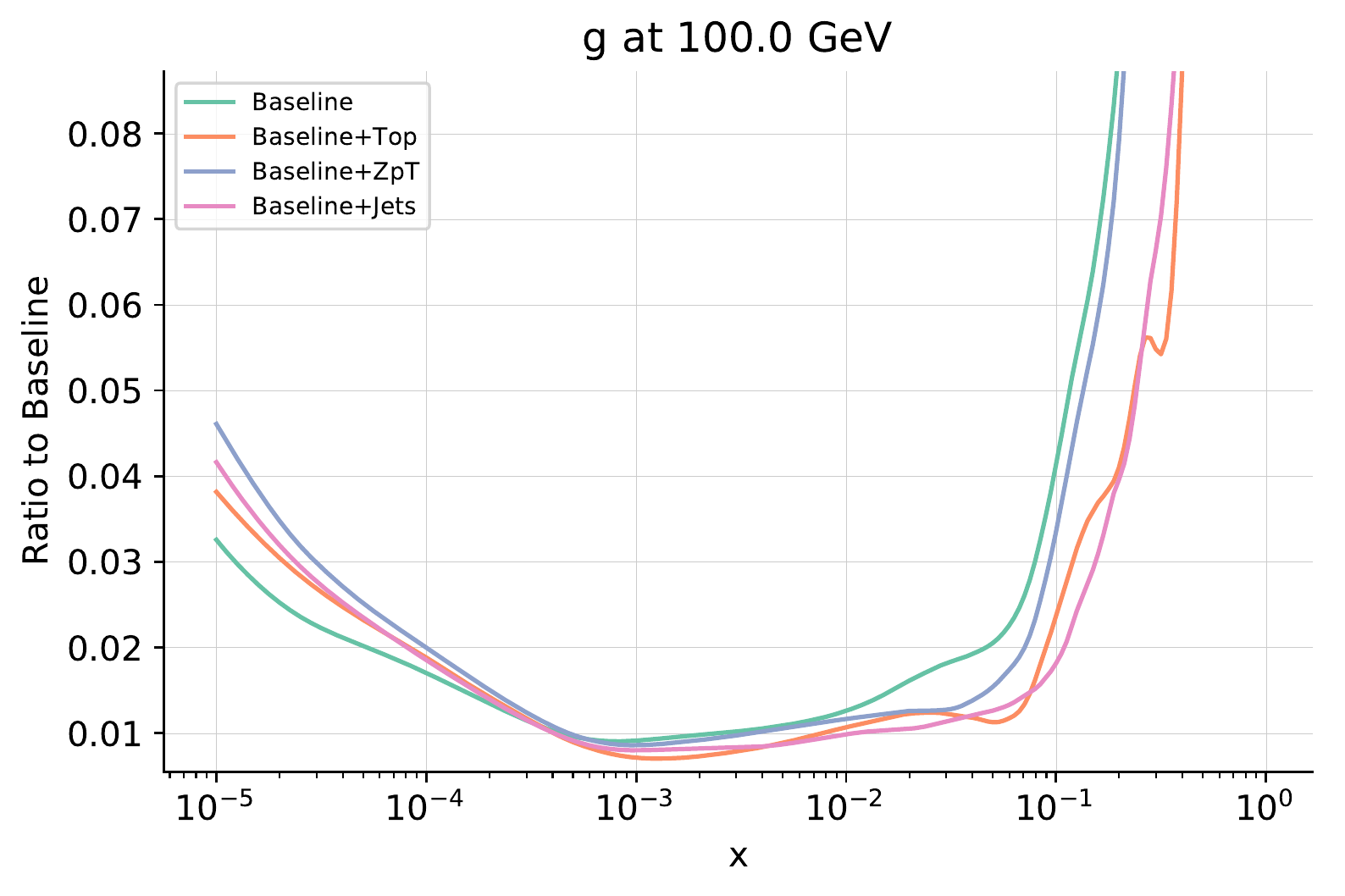}\\
\caption{Comparison between the gluon PDF as obtained from a baseline 
  fit that includes all the data in the NNDPF3.1 analysis except $Z$ $p_T$,
  top and jet measurements, and three fits in which each of these
  data sets is included in turn on top of the baseline data set:
  $Z$ $p_T$ measurements (first row), top differential distribution and total 
  cross section measurements (second row) and inclusive jet cross section
  measurements by using the exact NNLO $K$-factors (third row). 
  The results are displayed simultaneously in the fourth row.
  In the left panels, the gluon PDF is normalised to the baseline fit, in 
  the right panel relative uncertainties are shown.
  The factorisation scale is set to $Q=100$ GeV.}
\label{fig:comparison} 
\end{figure}

In NNPDF3.1, the three classes of measurements (inclusive jets, $Z$ 
$p_T$ and top data) have all been included at the same time in the global fit. 
This was compared to variants in which either of the three measurements was 
removed at a time from the global dataset.
This shows only indirectly the impact of each of these datasets.
A much more direct comparison is the one in which each dataset is added 
individually to a baseline dataset made up of all the data in NNPDF3.1 except 
the piece sensitive to the gluon at large $x$.
Therefore, we run four additional fits, with the same theoretical settings as
in the NNLO NNPDF3.1 default fit. 
In the baseline fit, all jet, $Z$ $p_T$ and top data are removed from the 
NNPDF3.1 NNLO fit.
In the other three fits, we add each of these measurements individually on top
of the baseline dataset.
The distance in the central value and uncertainty between each of these
three fits and the baseline is shown in Fig.~\ref{fig:largexdist}.
The corresponding gluon PDFs are compared in Fig.~\ref{fig:comparison}.

At the level of the central value, we observe from 
Figs.~\ref{fig:largexdist}-\ref{fig:comparison} that the three sets of 
observables favour a slightly softer gluon PDF then the baseline above
$x\sim 0.2$.
Such an effect, up to half a $\sigma$, is more pronounced for 
jet and top data.
At the level of ucnertainties, the transverse momentum distribution of the 
$Z$ boson decreases the gluon PDF uncertainty by almost a factor of two for 
$10^{-2}\lesssim x\lesssim 10^{-1}$, 
while keeping the central value well within the baseline error band. 
Top-pair and jet production data have a bigger impact, 
as the relative uncertainty is reduced by almost 100\% for all 
$x\gtrsim10^{-2}$. 
All data sets consistently pull the gluon central value in the same direction;
most notewordly, top data provide a constraint on it competitive with inclusive 
jet data.
 

In summary, the combined effect of $Z$ $p_T$, top and jet data is to 
consistently constrain the gluon PDF at medium-to-large values of $x$
with unprecedented, few percent, precision. 
The unprecedented level of precision in the
knowledge of the large-$x$ gluon is a remarkable achievement of the
LHC experimental program, which can only further improve thanks to 
data at higher luminosity and centre-of-mass energy in the future.

We are grateful to our colleagues in the NNPDF Collaboration, especially to 
S.~ Carrazza, S.~Forte, J.~Rojo and L.~Rottoli. 
E.R.N. is supported by the the STFC grant ST/M003787/1. 
M.U. is supported by a Royal Society Dorothy Hodgkin Research Fellowship and
partially by the STFC grant ST/L000385/1.
Plots in Figs.~\ref{fig:jetbins} and~\ref{fig:comparison} were drawn using the 
code of Ref.~\cite{zahari_kassabov_2019_2571601}.


\end{document}